\documentclass[twocolumn]{aastex6}
\usepackage{natbib}
\usepackage{graphicx}

\shorttitle{Observational Constraints on First-Star Nucleosynthesis. I. }
\shortauthors{Yoon et al.}

\begin{document}

\title{Observational Constraints on First-Star Nucleosynthesis. \\I.
Evidence for Multiple Progenitors of CEMP-no Stars}
\iffalse
\author{Jinmi Yoon\altaffilmark{1}, Timothy C. Beers, Vinicius M. Placco, \\Kaitlin C. Rasmussen, and Daniela Carollo}
\affil{Department of Physics, University of Notre Dame, Notre Dame, IN 46556, USA\\
Joint Institute for Nuclear Astrophysics and Center for the Evolution of the Elements (JINA-CEE), Notre Dame, USA}

\author{Siyu He}
\affil{Department of Physics, Xi'an Jiaotong University, \\Shaanxi, 710049, People's Republic of China}
\author{Terese T. Hansen}
\affil{Observatories of the Carnegie Institution of Washington, \\813 Santa Barbara St., Pasadena, CA 91101, USA}

\author{Ian U. Roederer}
\affil{Department of Astronomy, University of Michigan, 1085 S. University Avenue, Ann Arbor, MI 48109, USA\\
Joint Institute for Nuclear Astrophysics and Center for the Evolution of the Elements (JINA-CEE), USA}

\and 

\author{Jeff Zeanah}
\affil{Z Solutions, Inc., 9430 Huntcliff Trace, Atlanta, GA
30350,USA}

\altaffiltext{1}{jinmi.yoon@nd.edu}
\fi

\author{Jinmi Yoon\altaffilmark{1,2}, Timothy C. Beers\altaffilmark{1,2}, Vinicius M. Placco\altaffilmark{1,2}, Kaitlin C. Rasmussen\altaffilmark{1,2}, \\
Daniela Carollo\altaffilmark{1,2}, Siyu He\altaffilmark{3}, Terese T.
Hansen\altaffilmark{4}, Ian U. Roederer\altaffilmark{2,5}, and Jeff Zeanah\altaffilmark{6}}

\altaffiltext{1}{Department of Physics, University of Notre Dame, Notre Dame, IN 46556, USA; jinmi.yoon@nd.edu}
\altaffiltext{2}{Joint Institute for Nuclear Astrophysics -- Center for the Evolution of the Elements (JINA-CEE), USA}
\altaffiltext{3}{Department of Physics, Xi'an Jiaotong University, Shaanxi, 710049, People's Republic of China}
\altaffiltext{4}{Observatories of the Carnegie Institution of Washington, 813 Santa Barbara St., Pasadena, CA 91101, USA}
\altaffiltext{5}{Department of Astronomy, University of Michigan, 1085 S. University Avenue, Ann Arbor, MI 48109, USA}
\altaffiltext{6}{Z Solutions, Inc., 9430 Huntcliff Trace, Atlanta, GA
30350, USA}

\begin{abstract}

We investigate anew the distribution of absolute carbon abundance,
$A$(C) $= \log\,\epsilon $(C), for carbon-enhanced metal-poor (CEMP)
stars in the halo of the Milky Way, based on high-resolution
spectroscopic data for a total sample of 305 CEMP stars. The sample
includes 147 CEMP-$s$ (and CEMP-r/s) stars, 127 CEMP-no stars, and 31
CEMP stars that are unclassified, based on the currently employed
[Ba/Fe] criterion.  We confirm previous claims that
the distribution of $A$(C) for CEMP stars is (at least) bimodal, with
newly determined peaks centered on $A$(C)$=7.96$ (the high-C region) and
$A$(C)$ =6.28$ (the low-C region).  A very high
fraction of CEMP-$s$ (and CEMP-r/s) stars belong to the high-C region,
while the great majority of CEMP-no stars reside in the low-C region.
However, there exists complexity in the morphology of the $A$(C)-[Fe/H]
space for the CEMP-no stars, a first indication that more than one class
of first-generation stellar progenitors may be required to account for
their observed abundances. The two groups of CEMP-no stars we identify
exhibit clearly different locations in the $A$(Na)-$A$(C)
and $A$(Mg)-$A$(C) spaces, also suggesting multiple progenitors. The clear distinction in
$A$(C) between the CEMP-$s$ (and CEMP-$r/s$) stars and the CEMP-no stars
appears to be {\it as successful}, and {\it likely more astrophysically
fundamental}, for the separation of these sub-classes as the previously
recommended criterion based on [Ba/Fe] (and [Ba/Eu]) abundance ratios.
This result opens the window for its application to present and future
large-scale low- and medium-resolution spectroscopic surveys.

\end{abstract}


\section{Introduction}\label{intro}

At low iron abundances relative to the Sun, a substantial fraction of
the stars in the halo of the Milky Way have been found to be greatly
enhanced in carbon, the so-called carbon-enhanced metal-poor (CEMP)
stars. \cite{beers2005} originally divided such stars into several
sub-classes, depending on the nature of their neutron-capture element
abundance ratios -- CEMP-$s$, CEMP-$r$, CEMP-$r/s$, and
CEMP-no\footnote{CEMP-$s$ : [C/Fe] $>$ +1.0, [Ba/Fe] $>$ +1.0, and
[Ba/Eu] $>$ +0.5\newline CEMP-$r$ : [C/Fe] $>$ +1.0 and [Eu/Fe]$>$ +1.0
\newline CEMP-$r/s$ : [C/Fe] $>$ +1.0 and 0.0 $<$ [Ba/Eu] $<$+0.5 
\newline CEMP-no : [C/Fe] $>$ +1.0 and [Ba/Fe] $<$ 0.0}. As
discussed by these authors, and many since, the observed differences in
the chemical signatures of the sub-classes of CEMP stars are thought to
be tied to differences in the astrophysical sites responsible for the
nucleosynthesis products they now incorporate in their atmospheres,
including elements produced by the very first generations of stars.

\subsection{The Origin of CEMP Stars}

In this paper we focus on the two most populous sub-classes of the
carbon-enhanced metal-poor stars, the CEMP-$s$ and CEMP-no stars. Based on both
extensive observational follow-up and theoretical modeling, the elemental
abundance pattern associated with the CEMP-$s$ stars (carbon enhancement
accompanied by strong over-abundances of neutron-capture elements
produced by the main $s$-process) is thought to arise from an {\it
extrinsic} process -- mass transfer to the presently observed star from
an evolved binary companion. This companion, the site where the
enhancement of carbon and the $s$-process elements originally took
place, is expected to have been a low- to intermediate-mass ($\sim$1 to
$\sim$4 $M_{\odot}$) asymptotic giant-branch (AGB) star, which has now
evolved to become a faint white dwarf \citep[e.g.,][]{suda2004, herwig2005,
lucatello2005, bisterzo2011,starkenburg2014,hansen2015a}.
The mass-transfer process itself has proven challenging to model, despite
extensive efforts in recent years (see, e.g., \citealt{abate2013,
abate2015c,abate2015b}, and references therein).

Accreted material can potentially be mixed into the atmosphere of the
presently observed companion by several processes (e.g.,
thermohaline mixing, levitation, etc., see \citealt{stancliffe2007,
stancliffe2008,stancliffe2010,matrozis2016}). Additional processing can
also occur towards the tip of the giant branch \citep{placco2013,
karakas2014,placco2014c}, all of which may complicate interpretation of
the observed elemental-abundance patterns.

Binary mass transfer is thought to play a role in the origin of the
CEMP-$r/s$ stars as well \citep[e.g.,][]{jonsell2006,lugaro2009,
bisterzo2011,herwig2011}, but the origin of the $r$-process-like
component of their abundance pattern remains unclear \citep{abate2016}.
It remains possible that yet another nucleosynthetic process, the
so-called $i$-process, may need to be invoked to account for their
observed abundance patterns \citep{dardelet2015,hampel2016}. For the
purpose of the present analysis, we group the CEMP-$r/s$ stars along
with the CEMP-$s$ stars.  

In contrast to the CEMP-$s$ and CEMP-$r/s$ stars, a number of lines of
observational evidence (including long-term radial-velocity monitoring;
see \citealt{hansen2016a}) indicate that the distinctive abundance
patterns of CEMP-no stars (carbon enhancement with a lack of
neutron-capture element over-abundances) arose from an {\it intrinsic}
process\footnote{We intend this term to indicate that the observed
elemental abundances on the surface of the star were present in the gas
from which the star first formed, and not (as it is also used) patterns
arising from internal processing in the star of material that is later
transported to the stellar surface.}. The inference is that the presently
observed CEMP-no stars are indeed bona-fide second-generation stars,
born in natal clouds polluted by massive first-generation stars. A
number of astrophysical sites for the progenitors of the CEMP-no stars
have been suggested. The so-called ``faint supernovae'' or
``mixing-and-fallback'' models \citep{umeda2003,umeda2005,nomoto2013,
tominaga2014} hold that the gas from which CEMP-no stars formed was
enriched by a supernova without sufficient explosion energy to release
its full complement of synthesized heavy elements (which fall back to
the nascent neutron star or black hole at its center), and only the
lighter elements (including C, N, O, and other light elements such as
Na, Mg, Al, and Si) are expelled. Another possibility, the so-called
``spinstar'' model \citep[e.g.,][]{meynet2006,meynet2010,chiappini2013}
proposes that the gas from which CEMP-no stars formed was enhanced in
carbon (as well as N and O) by the strong stellar winds expected to
arise from rapidly-rotating massive stars of ultra low metallicity. In
addition, \citet{heger2010} have considered possible progenitors of the
CEMP-no stars including the effects of rotation and mixing and fallback.
More recent modeling has suggested that spinstars may also be capable of
producing other light elements and some amount of first-peak
neutron-capture elements (such as Sr) and second-peak $s$-process
elements (such as Ba), and possibly even third-peak elements such as Pb,
depending on the degree of the internal mixing induced by the rapid
rotation~\citep{maeder2015,frischknecht2016}. 

Finally, we note that \citet{cooke2011a,cooke2012} have reported on
recently discovered high-redshift carbon-enhanced damped Lyman-$\alpha$
systems that exhibit elemental-abundance patterns which resemble those
expected from massive, carbon-producing first stars. These
authors speculated that the progenitors that produced these patterns
are the same as those responsible for those associated with CEMP-no
stars in the Galaxy.

\subsection{The High and Low Carbon Bands for CEMP Stars} 

\citet{spite2013} used literature abundance data for
$\sim 50$ CEMP main-sequence turnoff and dwarf stars, including both
CEMP-$s$ and CEMP-no stars, and plotted the absolute carbon abundance,
$A$(C) $= \log\,\epsilon$(C)\footnote{$A$(X) = $\log\,\epsilon$(X) =
$\log\,$($N_{\rm X}/N_{\rm H}$)+12, where X represents a given
element.}, as a function of metallicity, [Fe/H], for their sample. The
stars in their sample were specifically chosen to be in evolutionary
stages where alteration of their surface elemental abundances, due to
significant internal mixing, were not expected to have occurred. We
point out, however, that certain processes, such as thermohaline mixing,
occur almost immediately after mass-transfer events (in CEMP-$s$ stars;
R. Stancliffe, priv. comm.), so some mixing (dilution) may have occured
even in supposedly unevolved stars. 

Based on this sample, they claimed the existence of a clear bimodality
among the CEMP stars -- the stars in their sample with [Fe/H] $> -$3.0,
which are dominated by CEMP-$s$ stars, populate a high-C ``plateau" at
$A$(C) $\sim 8.25$, close to the Solar value of $A$(C). In contrast, the
stars with [Fe/H] $ < -$3.4, which are exclusively CEMP-no stars, reside
in a lower region (and possible plateau) at $A$(C) $\sim 6.5$. They
interpreted this behavior as the result of the different
carbon-production mechanisms for these sub-classes of stars --
mass-transfer from binary AGB companions in the case of CEMP-$s$ stars
and enrichment of the natal clouds of the CEMP-no stars by massive-star
nucleosynthesis.

\citet{bonifacio2015} confirmed and extended the claim by Spite et al. 
with a larger sample ($\sim$70) of unevolved main-sequence turnoff
and dwarf stars, along with a few lower red giant-branch (RGB) CEMP
stars with [Fe/H] $ > -3.5$. These authors found a clear separation of
the $A$(C) distribution, but commented that the individual distributions
of carbon abundance were quite wide, on the order of one dex. They
advocated for a similar explanation of this separation as in Spite et
al., based on different carbon-production mechanisms for the CEMP-$s$
and CEMP-no stars.

The work of \citet{hansen2015a} provided new data for
additional CEMP stars, and considered them along with literature data (compilation from \citealt{yong2013}),
confirming once again the existence of the carbon bands, based on a 
total of 64 stars. However, they identified three CEMP-no stars located on
the high-C band, as well as the apparent existence of a smooth
transition of $A$(C) between the two bands, which as they noted presents
a challenge to the interpretation of the bimodality in $A$(C) as
exclusively due to extrinsic (AGB mass-transfer) and intrinsic
(C-enriched ISM) processes. These authors emphasized the crucial role
that knowledge of the binary status for stars associated with the two
carbon bands may play for determination of the nature of their
progenitors, and recently published the results of long-term
radial-velocity monitoring for samples of CEMP-no \citep{hansen2016a}
and CEMP-$s$ \citep{hansen2016b} stars.   

In order to further explore these questions, we have compiled an
extensive set of 305 CEMP stars with available high-resolution
spectroscopic data from the literature, including more recent data than
was available to the studies conducted in the past few years. 

This paper is arranged as follows. Section~\ref{data} describes details
of the literature data compilation, and the corrections we have applied
in order to place the data on a suitable common scale. The results of
our analysis, presented in Section~\ref{results}, clearly support the
existence of (at least) a bimodality in the distribution of $A$(C) for
CEMP stars, but we note that the $A$(C) distribution exhibits more
complex behavior that is {\it not captured} by its description as carbon
plateaus or bands. Instead, we suggest that the CEMP stars can be more
usefully described as falling into three groups, one for the CEMP-$s$
(and CEMP-$r/s$) stars and two for the CEMP-no stars, based on their
location in the $A$(C)-[Fe/H] space. We discuss these divisions in more
detail in Section~\ref{discussion}, and demonstrate the existence of a
correlated behavior between the absolute abundances of the light
elements Na and Mg, $A$(Na) and $A$(Mg), with $A$(C). Collectively, this
may provide the first evidence for the existence of at least two
distinct progenitor populations that are responsible for the abundance
signatures among CEMP-no stars. In this section we also consider
information that can be gleaned from the subset of CEMP stars with known
binary status, concluding that the carbon enhancement of the great
majority of CEMP-no stars is an intrinsic process, while most CEMP-$s$
(and CEMP-$r/s$) stars are extrinsically enriched, as previously
suggested. We also identify several interesting subsets of stars that
exhibit abundance anomalies relative to the majority of other CEMP stars
in our sample. Finally, we argue that classification based on $A$(C) is
likely to be a more astrophysically fundamental (and equally successful)
method to distinguish the CEMP-no stars from the CEMP-$s$ and CEMP-$r/s$
stars than the previously employed approach based on [Ba/Fe] (and
[Ba/Eu]) ratios, with the considerable advantage that it can be obtained
from low- to medium-resolution, rather than high-resolution,
spectroscopy. Our conclusions are briefly summarized in Section
\ref{conclusion}.

\section{Compilation of Literature Data} \label{data}

We have endeavored to compile as complete a list as possible of CEMP-$s$
(and CEMP-$r/s$) and CEMP-no stars having [Fe/H] $< -1.0$ and [C/Fe]
$\geq +$0.7 with available high-resolution spectroscopic abundance
information. We have only considered stars with claimed detections or
lower limits for carbon, along with several critical elemental abundance
ratios, such as [Ba/Fe] and [Eu/Fe]. The great majority of our sample
comes from the literature compilation of \citet{placco2014c}. To this,
we have added more recent literature data from a number of authors
\citep[e.g.,][]{roederer2014,hansen2015a,jacobson2015}, as well as a
number of more metal-rich CEMP stars (sometimes referred to as ``CH
stars'') with $-2.0 < $ [Fe/H] $\leq -1.0$, [C/Fe] $ \geq +0.7$, and
[Ba/Fe] $ \geq +1.0$. Our sample of these more metal-rich CEMP stars is
certain to be incomplete. For stars with multiple reported observations,
we have given preference to those stars with spectra having higher
resolving power and/or higher S/N; for those with similar quality
spectra, we kept the most recent data.
 
Because we compiled the literature data from studies conducted with a
variety of instruments, and analyzed with different atmospheric models
and methods of estimation, there are unavoidable inconsistencies in the
adopted abundance estimates. 
There are also differences between authors in the assignment of the
sub-classes for CEMP stars, in particular for the stars with 0.0 $<$
[Ba/Fe] $< +1.0$. Below we describe our attempt to resolve at least some
of these difficulties.

High-resolution spectroscopic analyses of metal-poor stars
conventionally estimate effective temperatures using broadband
photometric colors (e.g., $B-V$ or $J-K$), however, some studies
\citep[e.g., ][]{roederer2014} have chosen to estimate T$_{\rm eff}$
from the spectral lines themselves. It has long been recognized that
there exist systematic offsets between effective temperatures estimated
by these two methods \citep[e.g.,][]{frebel2013}. When available,
we collected the reported $A$(C)$=\log \epsilon$(C) from each study. 
For some studies, e.g., \citet{cohen2013}, as well as the compilation 
of \citet{placco2014c}, only [C/Fe] estimates were reported.  In these
cases, estimates of absolute carbon abundances, $A$(C), were 
derived using the relation $A$(C) = [C/Fe] $+$ [Fe/H] $+
A$(C)$_\odot$.
The systematic errors in the absolute abundance of $A$(C) introduced by
this, for T$_{\rm eff}$ differences up to 200-250~K, are within the
typical observational errors for $A$(C) $\sim$ 0.20-0.25 dex; we make no
explicit correction for their presence. However, these do not have
a major effect on derived elemental abundance ratios, [X/Fe], which
are similarly affected by temperature variations.\footnote{The typical
systematic error of $A$(Fe), due to a temperature variation of
$\sim$200-250~K is, in general, comparable to those of $A$(C). The
net effects on the abundance ratios are not significant when
compared to the typical total errors, $\sim$0.2-0.25 dex, in [C/Fe]. }

The overall stellar metallicity, [Fe/H], is often represented by
\ion{Fe}{2} lines, when feasible, because they represent the dominant ionization
state for the temperature range of F-G-K stars. However, for stars with
metallicities with [Fe/H] $ < -3.0$, the \ion{Fe}{2} lines are rarely
sufficiently numerous (or even detectable), hence many studies have been
forced to use neutral iron lines. Nevertheless, the \citet{roederer2014}
study was able to derive metallicities based on the \ion{Fe}{2} lines,
owing to the relatively bright stars in their sample. In addition, when
they calculated the individual element abundance ratios they used the
same ionization state of iron as for the element under consideration --
for instance, [\ion{Ba}{2}/\ion{Fe}{2}] for the Ba/Fe abundance ratio.
In order to be consistent with the majority of studies in our
compilation, we have recalculated [Fe/H] and the relative elemental
abundance ratios ([X/Fe]) based on \ion{Fe}{1} for this study. We have
also only considered results based on 1-D LTE assumptions. All
metallicities and abundances were re-scaled to the \citet{asplund2009}
Solar photospheric abundances. 

Evolved stars are known to have some degree of depletion in their
surface carbon abundances due to CN processing that takes place 
both because of first dredge-up and additional mixing on the
upper RGB \citep{gratton2000}. Therefore, for such stars, we have
estimated the ``original'' carbon abundances based on each star's
evolutionary state using an online carbon-correction
calculator\footnote{\href{http://www.nd.edu/~vplacco/carbon-cor.html}{http://www.nd.edu/$\sim$vplacco/carbon-cor.html}}.
For the stars added to our compilation since \citet{placco2014c}, and
provided without classifications by the original authors, we have
assigned CEMP sub-classes based on the definitions of \citet{beers2005},
but using [C/Fe] $ \geq +0.7$ \citep{aoki2007} as the criterion for carbon
enhancement. We note that some studies \citep[e.g.,][]{masseron2010,
spite2013, bonifacio2015, hansen2016b} classified CEMP sub-classes
somewhat differently for the CEMP stars with 0.0 $<$ [Ba/Fe] $< +$1.0;
we adopted their claimed classification.

\begin{figure*}[th!]
\includegraphics[]{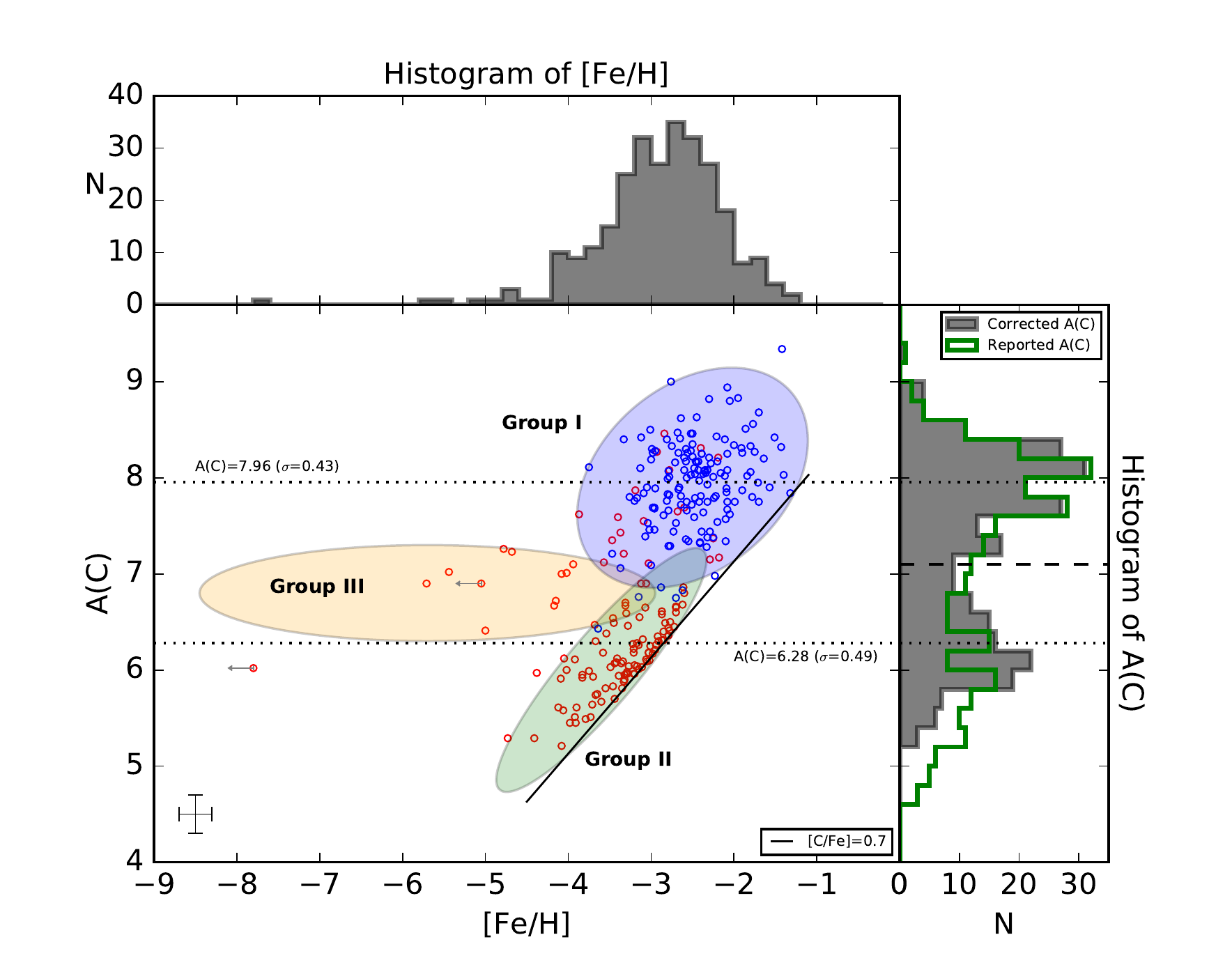}
\caption{\footnotesize Scatter diagram of the corrected $A$(C) vs. [Fe/H] for our
compilation of CEMP stars. The blue and red open circles represent the
147 CEMP-$s/rs$ stars and 127 CEMP-no stars, respectively (the 31
unclassified CEMP stars are not shown). The black dotted lines indicate
the estimated locations of the carbon peaks, based on a two-component
Gaussian fit to the corrected $A$(C) distribution. The majority of the
CEMP-$s/rs$ stars reside in a region surrounding the high-C peak at
$A$(C) $\sim$8.0, while the majority of CEMP-no stars scatter around the
low-C peak at $A$(C) $\sim$6.3. The black solid line provides
a reference at [C/Fe] = $+0.7$. The gray shaded histogram in the
top margin shows the metallicity distribution of the full sample. The
gray shaded histogram in the right margin is the corrected $A$(C)
distribution; the green unfilled histogram is the ``as reported'' $A$(C)
distribution. Note that the 31 unclassified CEMP stars are not
included in these fits. The black dashed line in the marginal histogram
of $A$(C) represents the midpoint of the two $A$(C) peaks, used for
separation of CEMP-$s/rs$ stars from CEMP-no stars, as described in the
text. A typical error bar for the sample stars we consider is shown at
the bottom left.
\label{fscatter}}
\end{figure*}

Our full sample consists of 305 CEMP stars: 147 CEMP-$s/rs$ stars
(hereafter, we employ this notation to include both the CEMP-$s$ and
CEMP-$r/s$ stars), 127 CEMP-no stars, and 31 CEMP stars that are not
sub-classified by the conventional criteria. Table~\ref{tbl-1} lists the
stellar parameter estimates for our program sample in columns (2)-(4).
Column (5) lists the reported carbon abundance ratio, [C/Fe]; column (6)
is the corrected carbon abundance ratio, [C/Fe]$_c$; and column (7) is
the absolute carbon abundance, $A$(C), based on the corrected carbon
abundance ratio. Columns (8) and (9) are the reported [Ba/Fe] and
[Eu/Fe] ratios, respectively, corrected in some cases as described
above. Below we discuss an alternative classification scheme for CEMP
stars, based on $A$(C), rather than one that requires use of the [Ba/Fe]
ratio. Column (10) of Table~\ref{tbl-1}, labeled as ``Class (Ba$\mid
A$(C))", lists the sub-classification of a given star based first on the
[Ba/Fe] criterion, followed by its classification based on the $A$(C)
criterion described below. For instance, the notation (no $\mid s$)
means that the star was sub-classified as a CEMP-no star based on the
[Ba/Fe] criterion, but as a CEMP-$s/rs$ star based on the $A$(C)
criterion. We note that precise sub-classification for these stars (as
either CEMP-$s$ or CEMP-$r/s$) requires a [Eu/Fe] detection, which is
not presently available for all of the stars in our sample. The binary
status of a given star, if known, is listed in column (11). We also
indicate cases of stars with disparate $A$(C) abundances and [Ba/Fe]
abundance ratios (for instance, high $A$(C) with low [Ba/Fe]), including
stars with anomalous $A$(C) or [Ba/Fe] given their known binary status.
We marked these objects with a $\surd$ in column (12) as ``Interesting
outliers'' \footnote{We group these stars along with other outliers having similar
behavior in Table~\ref{tbl-3}, as described below.}. Column (13) of the
table provides the reference to the original study from which our
stellar parameter and abundance information was drawn.

For convenience of the subsequent analysis, we have listed our program
stars in Table~\ref{tbl-1} in the groupings described in
Section~\ref{results} below (Group~I, Group~II, and Group~III, along with a
group without sub-classifications based on [Ba/Fe]).\\

\begin{figure*}
\begin{center}
\plottwo{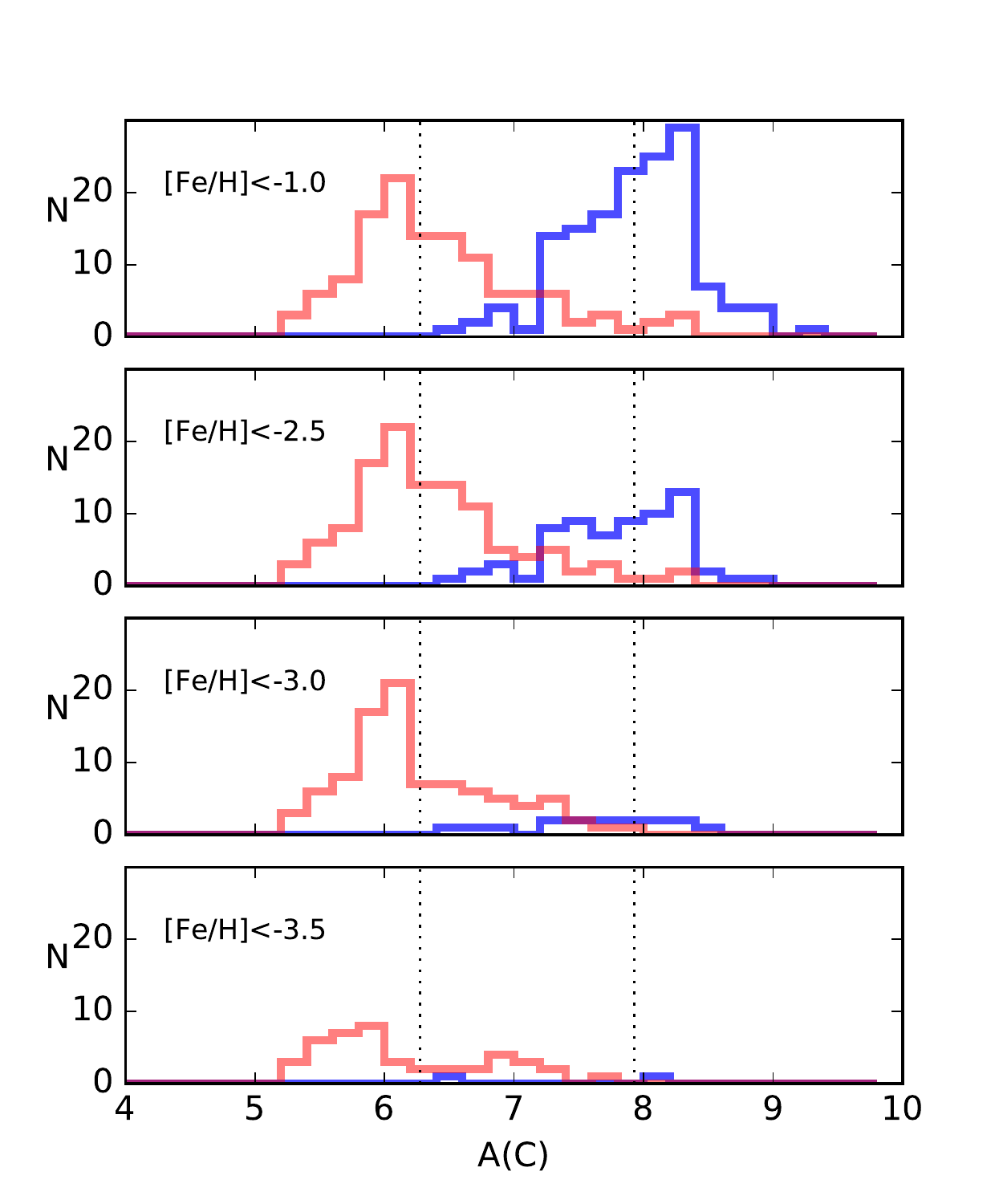}{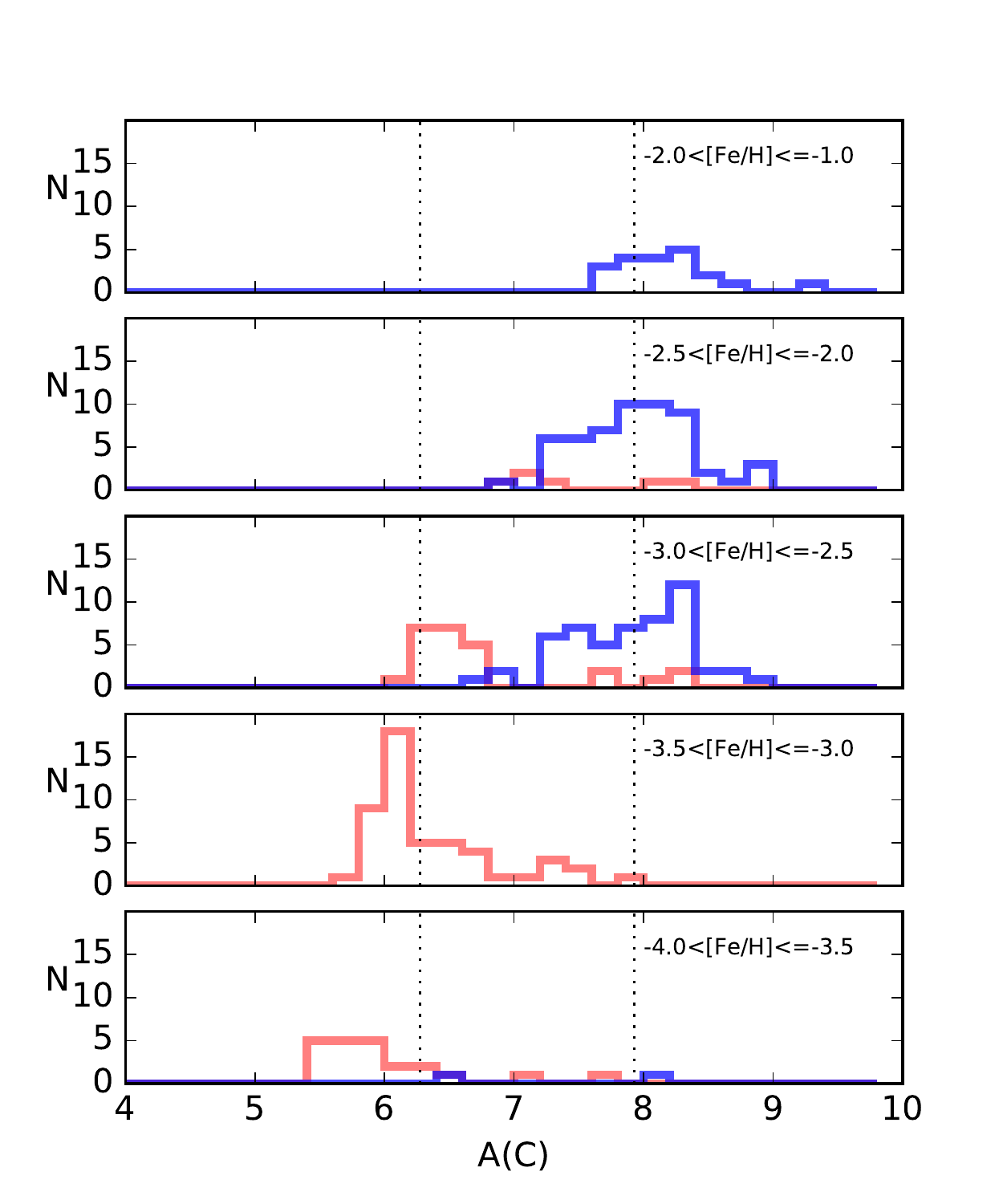}
\caption{\footnotesize Left panel: Cumulative histograms of $A$(C) for the  
the CEMP-$s/rs$ stars (blue) and CEMP-no stars (red), in cuts of
decreasing metallicity. Right panel: Differential histograms of $A$(C)
for both sub-classes in intervals of declining metallicities. The dotted
black lines represent the estimated location of the high-C and low-C
peaks based on a two-component Gaussian fit (see text).
\label{fhistogram}}
\end{center}
\end{figure*}

\section{Results}\label{results}

Figure~\ref{fscatter} shows the corrected $A$(C) distribution, as a
function of [Fe/H], for our compiled sample of sub-classified CEMP (147
CEMP-s/rs and 127 CEMP-no) stars. The remaining 31 unclassified CEMP
stars are not included in Figure~\ref{fscatter} and subsequent figures,
due to their lack of available [Ba/Fe] measurements. The blue and
red open circles represent the CEMP-$s/rs$ and CEMP-no stars,
respectively. Note from the marginal histogram shown on the right side
of the figure that a significant number of the stars among the CEMP-no
sub-sample (those with the lowest $A$(C), which generally correspond to
the more evolved, higher luminosity giants that undergo CN processing)
have $A$(C) values that have been revised upward relative to their ``as
reported'' values, whereas the CEMP-$s/rs$ stars have only a small
number of corrections applied (see \citealt{placco2014c}). The
bimodality of the $A$(C) distribution is much clearer post correction,
underscoring the importance of carrying out this step.

Although the bimodality of $A(C)$ is clear in Figure~\ref{fscatter}, the
peaks of the distribution are about 0.3\,dex lower than the
determinations by \citet{spite2013}, who claimed $A$(C) $\sim 8.25$ and
$A$(C) $\sim 6.5$ for the CEMP-$s$ and CEMP-no stars,
respectively\footnote{It should be recalled that \citet{spite2013}
considered only the ``unmixed'' CEMP stars in their discussion, whereas,
in our attempt to build a larger total sample, we have included both
unevolved and evolved (giant) stars. When only unevolved stars
($\sim$120 stars with $\log$g $ > $ 2.5) in our sample are taken into account for the
peak estimates, the low-C peak is located at A(C) $\sim$ 6.6, similar to
the location reported by Spite et al.}. We obtained
estimates of the peaks in our distribution by fitting a two-component
Gaussian distribution~\citep{scikit} to the corrected $A$(C) values,
obtaining peak values of $A$(C) = 7.96 (with a dispersion of 0.43 dex)
for the high-C region, and $A$(C) = 6.28 (with a dispersion of 0.49 dex)
for the low-C region, respectively. These peaks are indicated by the
black dotted lines in the figure; the black dashed line shown in the
marginal histogram of $A$(C) is located at the midpoint between these
peaks, at $A$(C) = 7.1. The previous claim that most CEMP-$s/rs$ stars
are associated with the high-C region, while the majority of CEMP-no
stars are associated with the low-C region, is clearly supported. 

There are remarkable differences in the morphology of the $A$(C) vs.
[Fe/H] distributions between the CEMP-$s/rs$ stars and the CEMP-no
stars, as seen from inspection of Figure~\ref{fscatter}. The
distribution of $A$(C) for the CEMP-$s/rs$ stars, indicated with the
blue shaded ellipse in Figure~\ref{fscatter}, exhibits a very weak
dependence on [Fe/H] -- there exists a wide scatter of $A$(C) values for
these stars at any given [Fe/H]. We refer to this subset of stars in the
$A$(C) vs. [Fe/H] diagram as ``Group I" stars. In contrast, the CEMP-no
stars exhibit two very different behaviors in the $A$(C) vs. [Fe/H]
space. For convenience, we refer to these as ``Group~II" and
``Group~III" stars\footnote{ We note that application of objective
clustering procedures yielded similar results. Three elliptical clusters
were determined using the mclust routine \citep{fraley2002,fraley2012},
implemented in R using the EM Mixed Model algorithm and the Bayesian
Information Criterion (BIC). The algorithm was constrained to select
three clusters. If left unconstrained, the algorithm would further
subdivide the data points of Group~II, however, the apparent
sub-structure in this group is likely introduced by selection effects in
the observations made to date, hence it was suppressed.} and indicate
them in the figure with the green and orange shaded ellipses,
respectively. Note that all three groups are defined based on their
morphology in the $A$(C) vs. [Fe/H] space alone; each group contains a
small fraction of CEMP stars with sub-classifications that differ from
the majority within the group -- these are discussed in more detail
below. The stars are listed by these different morphological groups in
Table~\ref{tbl-1}.\footnote{The stars in the overlapping regions are
also listed as such in the table notes.}

The Group~III stars are located in the lowest metallicity regime, [Fe/H]
$< -$3.5, centered on $A$(C) $\sim$ 6.8, higher than the $A$(C) values
for many of the Group~II stars. The $A$(C) values for the Group~III
stars also exhibit no clear dependence on [Fe/H]. We note that the
lowest [Fe/H] star in our sample (SMSS~J0313-6708, with [Fe/H] $ <
-7.8$) has a lower $A$(C) (by about 0.3 dex) than the rest of the
Group~III stars. We assign this star to Group~III, as its location is
more similar to other Group~III stars than to the Group~II stars. 

The Group~II stars are clustered in the metallicity range $-5.0
\la$ [Fe/H] $\la -2.5$, and, in contrast to the Group~III stars,
exhibit a clear dependence of $A$(C) on [Fe/H]. As seen in the figure,
there are also CEMP-no stars that fall into neither of these groups, but
are scattered throughout the region primarily occupied by the
CEMP-$s/rs$ stars -- these stars are placed into Group~I.

There also exists a contrast in the metallicity distributions between
the CEMP-$s/rs$ and CEMP-no stars, as seen in Figure~\ref{fscatter}. The
CEMP-$s/rs$ stars in our sample cover a range of $\sim$ 2.5 dex in
metallicity, with the mean at [Fe/H] $= -2.47$ (recall that our sample
is incomplete at the metal-rich end, so the mean value is likely
somewhat higher in reality). By comparison, the CEMP-no stars are spread
over $\sim$ 6 dex in metallicity, with the mean at [Fe/H] $= -3.42$. This
contrast is seen in Figure~\ref{fhistogram} as well. The left panels are
cumulative histograms of $A$(C) for each CEMP sub-class over different
cuts of decreasing [Fe/H]. The right panels are differential histograms
of $A$(C) for the CEMP sub-classes in specific metallicity ranges. As
claimed in \citet{spite2013}, most of the CEMP-$s/rs$ stars reside in
the high-C region, while the CEMP-no stars predominantly belong to the
low-C region, as shown in the top left panel, although they are
asymmetrically distributed, with a long tail toward higher $A$(C) in all
the metallicity cuts. As noted by previous work \citep[e.g.,
][]{aoki2007, placco2014c}, the CEMP-no stars are dominant in the
metallicity range [Fe/H] $< -3.0$, while most CEMP-$s/rs$ stars are
found above this metallicity, consistent with the behavior of both the
cumulative and differential histograms shown in Figure~\ref{fhistogram}.
We note that there are two CEMP-$s$ stars in our sample with quite low
metallicities compared to the rest of the CEMP-$s/rs$ stars,
CS~22960-053 with [Fe/H] = $-3.64$~\citep{roederer2014} and HE~0002-1037
with [Fe/H] = $-3.75$~\citep{hansen2016b}. 

\section{Discussion}\label{discussion}

\begin{figure*}[th!]
\begin{center}
\includegraphics[scale=.8]{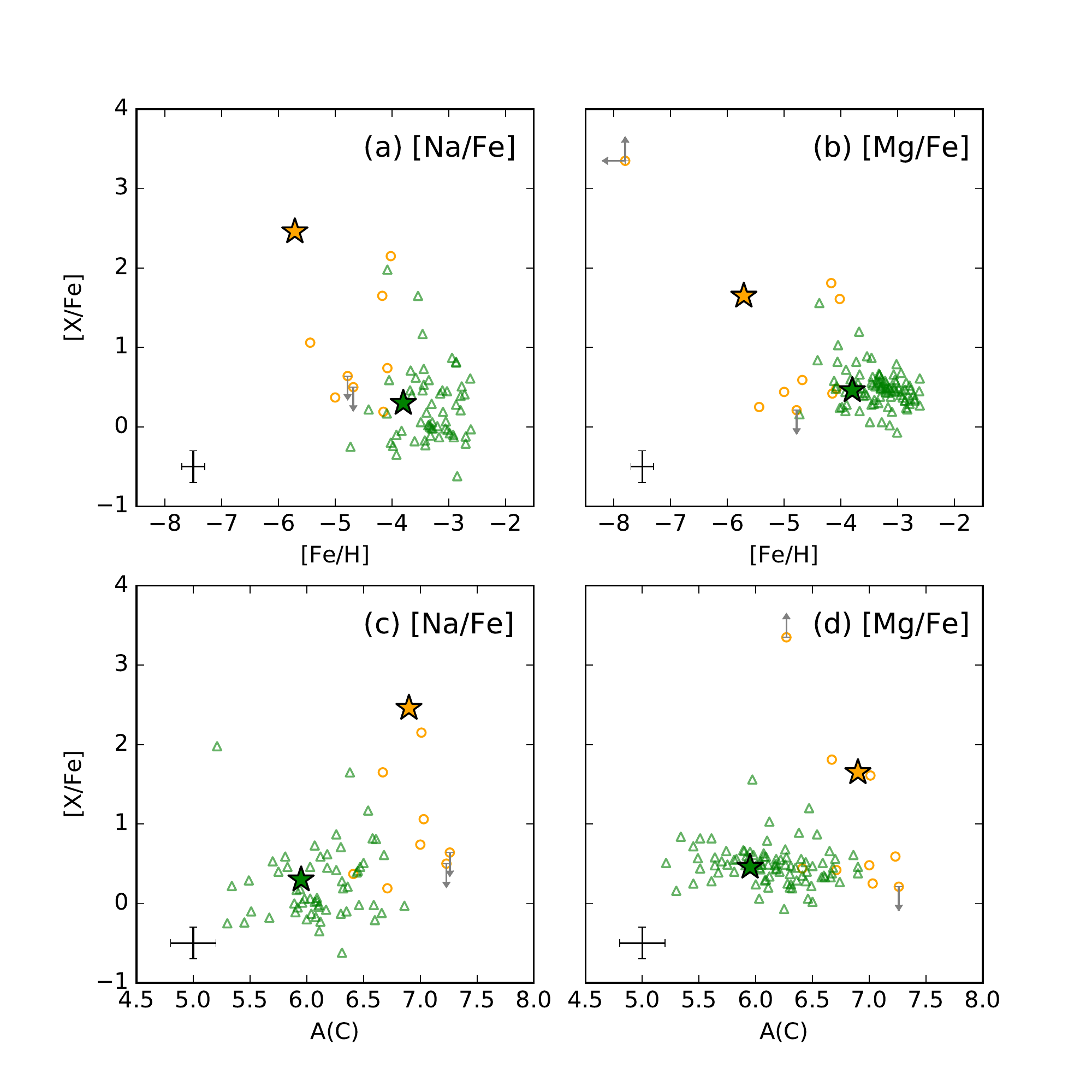}
\caption{Panels (a) and (b):  Distribution of the abundance ratios 
[Na/Fe] and [Mg/Fe], as a function of [Fe/H]. Panels (c) and (d): The
same ratios, as a function of $A$(C). The open green triangles and
orange circles represent Group~II and Group~III stars, respectively. The
filled green and orange stars are the well-studied CEMP-no stars BD+44
493 and HE~1327-2326, which we take as type examples of Group~II and
Group~III CEMP-no stars, respectively. Note that only stars with
available classifications, based on the [Ba/Fe] criterion, are shown. A
typical observational error bar is shown in the bottom left of each
panel.
\label{fna-mg-ratio}}
\end{center}
\end{figure*}

In this section we consider the astrophysical implications of the
observed $A$(C) vs. [Fe/H] distributions for CEMP stars, and speculate
on what more might be revealed from further detailed studies. We also
summarize the known binary status of our sample of CEMP stars, and
highlight several cases of ``interesting outliers" -- stars whose binary
status does not meet with expectation given their observed $A$(C) and/or
[Ba/Fe], or with disparate values of $A$(C) and [Ba/Fe]. Finally, we
consider whether $A$(C) can be used as an effective discriminator
between the dominant populations of CEMP stars, as an alternative to the
[Ba/Fe]-based scheme for CEMP sub-classification that is in conventional
use.

\subsection{The Complex Behaviors in the $A$(C) vs. [Fe/H] Distribution}

The different behaviors shown in the $A$(C) distributions for the
CEMP-$s/rs$ and CEMP-no stars require further detailed investigation,
and larger samples are clearly desirable. However, on its face, it
appears that reference to carbon plateaus or bands is no longer a valid
description of the observations -- the morphology of the $A$(C)-[Fe/H]
space is much more rich and complex. In retrospect, it is perhaps not
surprising that the observed $A$(C) associated with mass-transfer
objects, such as the CEMP-$s/rs$ stars, and that associated with the
(presumably) high-mass stellar progenitors of CEMP-no stars should
differ so dramatically from one another, as they arise from greatly
contrasting nucleosynthetic pathways. Even among the CEMP-no stars
themselves, the rather striking differences in $A$(C) vs. [Fe/H] for the
Group~II and Group~III stars calls out for an astrophysical
interpretation. It seems plausible that different classes of
progenitors, such as the proposed faint mixing-and-fallback supernovae
and/or massive ultra low-metallicity spinstars, or other yet-to-be
suggested sites, may be able to account for the contrasting behaviors
that are observed.

The $A(C)$ distribution of the CEMP-$s/rs$ stars in Group I suggests
a single class of progenitors, with no obvious dependence on
[Fe/H], based on data presently in hand. We note that the $A$(C)
distribution of the CEMP-$r/s$ stars exhibits no clear difference from
that of the CEMP-$s$ stars. Better understanding of the underlying
processes and their astrophysical implications requires detailed
analyses of the full abundance patterns for these stars, more complete
knowledge of their binary status (and orbital parameters), and, in
particular, more theoretical population-synthesis modeling along the
lines of \citet{abate2015a}. We defer a more thorough discussion of the
abundance patterns of these stars to future work.

The CEMP-no stars of Groups II and III present several compelling
behaviors. The clear difference in the $A$(C) vs. [Fe/H] distributions
for these two groups has been described above. Here, we examine whether
contrasts in the observed abundances for other elements might exist.   

It has been previously noticed that a substantial fraction of the
extremely metal-poor stars with [Fe/H] $< -$3.0 exhibit moderate to
strong enhancements in their light-element abundance ratios, such as
[Na/Fe], [Mg/Fe], [Al/Fe], and [Si/Fe]; this ``light-element signature''
appears to be ubiquitous among the lowest metallicity stars
\citep[e.g.,][]{aoki2002c,christlieb2004,frebel2005,aoki2006,frebel2008,
norris2013}. We have examined the behavior of the two most commonly
reported light elements, Na and Mg, for the CEMP-no stars in Groups II
and III with this information available, listed in Table~\ref{tbl-2}. 

\begin{figure*}[th!]
\begin{center}
\includegraphics[scale=.8]{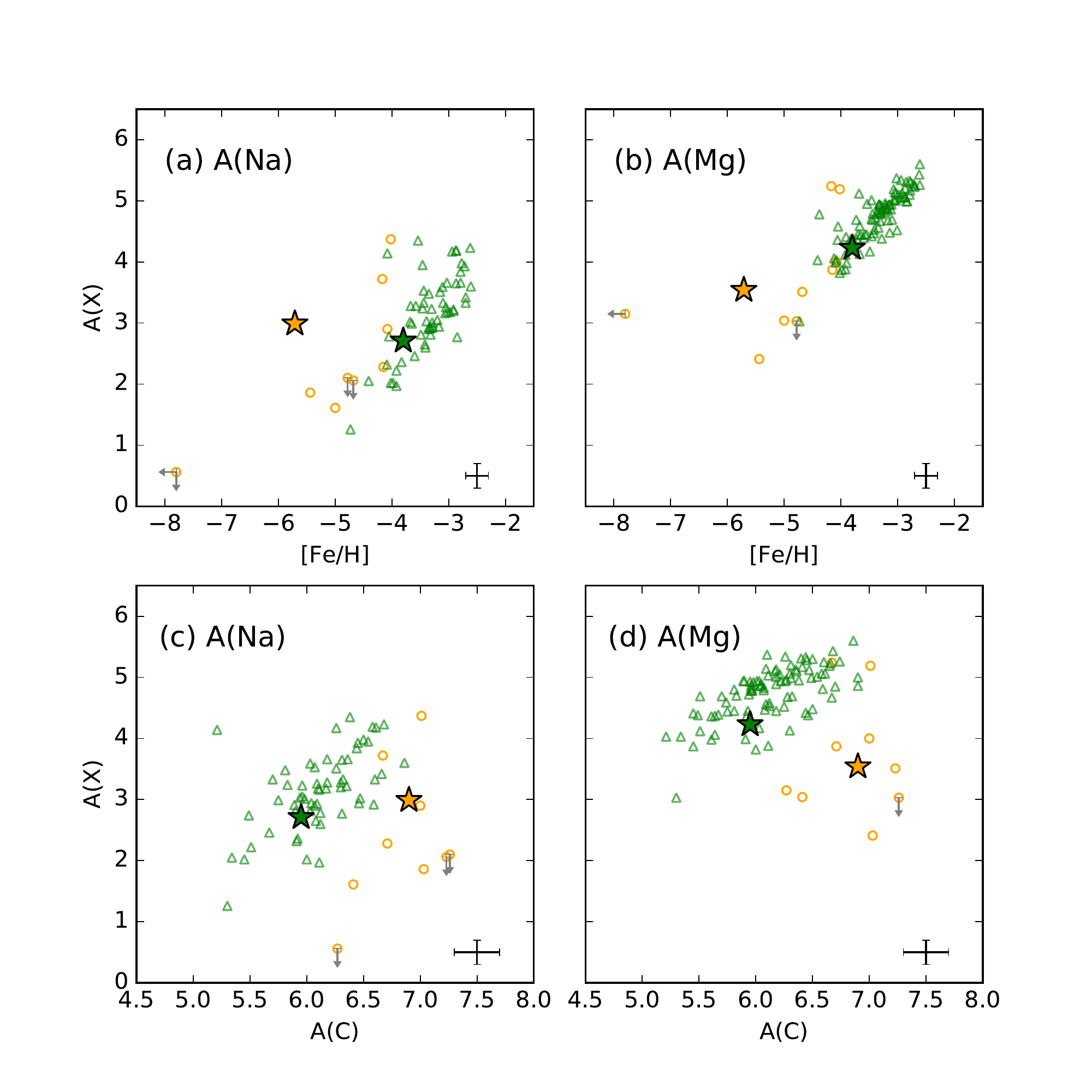}
\caption{Panels (a) and (b): Distribution of the absolute abundances of Na and Mg, $A$(Na) and $A$(Mg), as a function of [Fe/H]. Panels (c) and (d): The same absolute
abundances, as a function of $A$(C). The colors and symbols are the same
as in Figure~\ref{fna-mg-ratio}. Note that only stars with available
classifications, based on the [Ba/Fe] criterion, are shown. A typical
observational error bar is shown in the bottom right of each panel.
\label{fna-mg}}
\end{center}
\end{figure*}

The distributions of the abundance ratios [Na/Fe] and [Mg/Fe] for the
Group~II stars (green open triangles) and Group~III stars (orange open
circles), as functions of [Fe/H] and $A$(C), are shown in
Figure~\ref{fna-mg-ratio}. This figure highlights the location of two
stars, BD+44 493 and HE~1327-2326, as filled green and orange stars,
respectively. We consider these stars as canonical examples of Group~II
and III CEMP-no stars, respectively. The previously noted presence of
enhancement in these ratios, with respect to [Fe/H], is apparent in the
upper panels in Figure~\ref{fna-mg-ratio}. Many of the stars with [Fe/H]
$< -4.0$ exhibit over-abundances of Na and Mg, although some do not. The
largest enhancements are found for the Group~III stars, although there
are also a number of Group~II stars with this signature present as well.
The lower panels show these same ratios with respect to $A$(C).
Enhancements in [Na/Fe] and [Mg/Fe] for stars with $A$(C) $\ga 6.5$ can
be seen, where once again the largest over-abundances are found for the
Group~III stars, along with a few exceptional Group~II stars.

Figure~\ref{fna-mg} shows the distributions of absolute abundances,
$A$(Na) and $A$(Mg), as functions of [Fe/H] and $A$(C). There is no
clear separation of the absolute abundances of Na and Mg for Group~II
and Group~III stars seen in panels (a) and (b); both scale roughly with
[Fe/H]. However, when these same absolute abundances are plotted vs.
$A$(C), shown in panels (c) and (d), an apparent dichotomy emerges --
the Group~II stars scale roughly with $A$(C), while the Group~III stars
exhibit no correlation with $A$(C), but are instead offset to generally
lower values for $A$(C) $\ga 6.0$, with only a few exceptions. Based on
work in progress, we note that there exist similar behaviors in the
$A$(Al)-$A$(C) and $A$(Si)-$A$(C) spaces as well. The two Group~III
stars with relatively high $A$(Na) and $A$(Mg) seen in the lower panels
(HE~1012-1540: [Fe/H] = $-$4.17, $A$(C) = 6.67, [Na/Fe] = $+1.65$,
[Mg/Fe] = +$1.81$, and HE~2139-5432: [Fe/H] = $-$4.02, $A$(C) = 7.01,
[Na/Fe]= $+2.15$, [Mg/Fe] = $+1.61$) are selected near the extremum of
the region identified with this group shown in Figure~\ref{fscatter},
close to the locations of the Group~II stars. These two stars also
present behaviors that are more similar to Group~II stars in all panels
of Figures~\ref{fna-mg-ratio} and~\ref{fna-mg}; their identification as
Group~III stars (which is somewhat arbitrary at this early stage of
understanding) is perhaps suspect. 

\begin{figure*}[hbt!]
\begin{center}
\includegraphics[scale=.6]{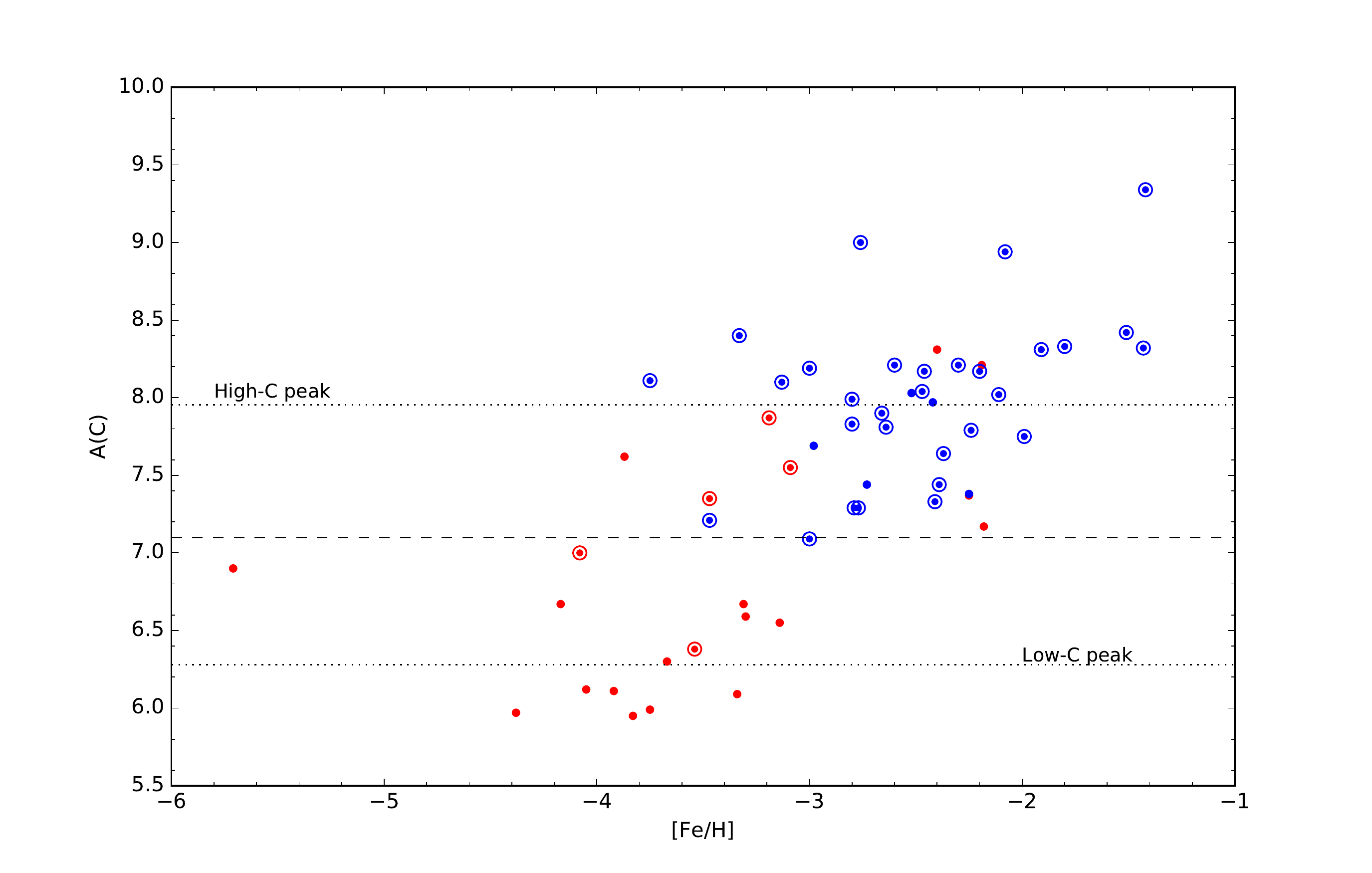}
\caption{The distribution of $A$(C), as a function of [Fe/H], of the known 
(and likely) binaries and known (and likely) single stars in our
compiled sample of CEMP stars. As before, the blue and red colors
indicate the CEMP-$s/rs$ and CEMP-no stars, respectively. The dots
($\cdot$) represent the known/likely single stars; the small dots inside
the larger open circles ($\sun$) represent the known/likely binary
stars. The dotted black lines indicate the location of the two $A$(C)
peaks described in the text. The dashed black line is a fiducial at
$A$(C) = 7.1, placed at the mid-point of the peaks in the marginal
histograms of $A$(C) shown in Figure~\ref{fscatter}; see text for
further discussion.\label{fbinarity}}
\end{center}
\end{figure*}

Based on the behaviors noted above, we argue that it is likely that more
than one class of progenitors for the CEMP-no stars of Groups~II and III
exist. To our knowledge, there is no clear predicted signature that
differentiates between the two most frequently considered progenitor
models for CEMP-no stars. However, the two canonical Group~II and
Group~III stars we have highlighted above have indeed been associated by
previous authors as type examples of the faint mixing-and-fallback
supernovae models (BD+44 493; e.g., \citealt{ito2013,placco2014b,
roederer2014, roederer2016}) and spinstar models (HE~1327-2326; e.g.,
\citealt{maeder2015,maeder2015b}). We note, however, that the observed elemental
abundance distribution of HE~1327-2326 has also been well-fit to a faint
supernova model \citep{Iwamoto2005,tominaga2007}, although the models
considered under-produce its observed [N/Fe] and do not predict its
[Sr/Fe] abundance. 

It may be premature to associate Group~II stars with faint
mixing-and-fallback SNe progenitors (15-40\,M$_\odot$;
\citealt{nomoto2013}) and the Group~III stars with spinstar progenitors
($\gtrsim $ 50\,M$_\odot$; \citealt{meynet2006}), and we remain
open to the possibility that other classes of progenitors may contribute
in the early chemical history of the Universe\footnote{Neutron-capture
abundance patterns found for a few of the CEMP-no Group~II stars
indicate a small amount of $r$-process material is present
\citep{roederer2014b}, which is not predicted to be produced by
mixing-and-fallback SNe.}.
Further theoretical development and modeling of the environments (e.g., the
amount of baryonic mass available for mixing and dilution of the yields
from both classes of model progenitors, expanding on the initial efforts
of, e.g., \citealt{cooke2014,susa2014}) is highly encouraged.

\subsection{Single Stars vs. Binary Stars}\label{binarity}

Knowledge of a given CEMP star's binary status is crucial 
for confident identification of its likely source of carbon enrichment,
whether extrinsic (AGB binary mass-transfer) or intrinsic (the star is
born in a previously C-enhanced environment).

Previous investigations have shown that a high fraction of the CEMP-$s$
stars are in binary systems \citep[e.g.,][and references
therein]{lucatello2005,starkenburg2014,jorissen2016}, prompting the
suggestion that $\it{all}$ CEMP-$s/rs$ stars have binary companions.
Most recently, \citet{hansen2016b} report that, although
82$\pm$26\% (18 of 22)\footnote{The published error bar on the CEMP-$s$
fraction in \citealt{hansen2016b} of 10\% is formally incorrect, and
should be substantially higher, 26\%, when calculated from traditional error propagation,
which we have used here.  However, it should also
be recognized that fractions of binary/non-binary stars for a given
population are in fact correlated, in the sense that their sum remains fixed, hence
improved statistical treatment of the appropriate error bars, rather than 
assigning errors from strictly Poisson statistics, needs to be considered
in the future.} of their sample
of CEMP-$s/rs$ stars are indeed binaries, there remain a small fraction
(18$\pm$10\%; 4 of 22) of stars that, despite a long-term campaign of
precision radial-velocity measurements, appear to be single (or binaries
with extremely long periods). As pointed out by these authors, the
existence of this handful of ``anomalous'' C- and Ba-enhanced single
stars opens the door to the possibility of enrichment by massive stars
capable of producing Ba (and perhaps other heavy elements) via a weak
$s$-process (see, e.g., \citealt{frischknecht2016}, and references
therein).

\citet{hansen2016a} also obtained precision long-term radial-velocity
monitoring observations of a sample of CEMP-no stars, reporting that the
binary fraction of CEMP-no stars is 17$\pm$9\% (4 of 24)\footnote{Three stars 
(CS~22166-016, CS~29527-015, and CS~22878-027)
were excluded in our calculations here, since the available
high-resolution data is not able to clearly classify them as CEMP
stars.}, i.e., no different than the observed binary fraction of
C-normal giants in the halo (16$\pm$4\%;
\citealt{carney2003}). It is clear that these two sub-classes of CEMP 
stars have different binary status, with implications for the nature of
their progenitors.

Figure~\ref{fbinarity} shows the distribution of $A$(C) as a function of
[Fe/H], for CEMP stars with known (or likely, meaning that further
confirmation is desired) binary status, including 35 CEMP-$s/rs$ stars
and 22 CEMP-no stars \citep{dearborn1986,mcclure1990, preston2001,
thompson2008, spite2013, placco2015b, hansen2016a,hansen2016b,
jorissen2016}. The dots in this figure represent single stars, while the
binary stars are shown as a dot inside a large open circle. The binary
status of these stars, where known, is also listed in Table~\ref{tbl-1}.
Although many CEMP stars do not yet have observational constraints on
their binary nature, the information in hand does provide the basis for
a number of interesting interpretations. 

Most CEMP-$s/rs$ stars (86$\pm$21\%; 30 of 35) shown in
Figure~\ref{fbinarity} are recognized binaries, and the majority (77$\pm$25\%; 17 of 22) 
of the CEMP-no stars are single stars. We note in
passing that it is remarkable how well the division of the stars by the
fiducial line at $A$(C) = 7.1, based on the marginal histogram of $A$(C)
from Figure~\ref{fscatter}, and discussed further below, effectively
separates the binary nature of {\it all} of the CEMP-$s/rs$ stars and
most of the CEMP-no stars. This clear contrast of the binary fraction
between the two sub-classes strongly supports the hypothesis that the
over-abundance in carbon and $s$-process elements for most CEMP-$s/rs$
stars is likely extrinsic, while the carbon enhancement in most CEMP-no
stars is the result of an intrinsic process. Therefore, the scatter in
$A$(C) among the CEMP-$s/rs$ stars shown in Figure~\ref{fscatter} might
be explained by mass-transfer models of AGB stars with differing masses,
mass-transfer efficiencies, and possible post-transfer dilution (e.g.,
\citealt{abate2015b}, and references therein). Further theoretical work
is required, of course, in order to establish if the observed
distribution of $A$(C) can be captured by extant models, or whether
additional complexity must be introduced. Regardless, the sample of
CEMP-$s/rs$ stars with measured $A$(C) is now sufficiently large that
such constraints should be feasible.  

\begin{figure*}
\begin{center}
\includegraphics[scale=0.8]{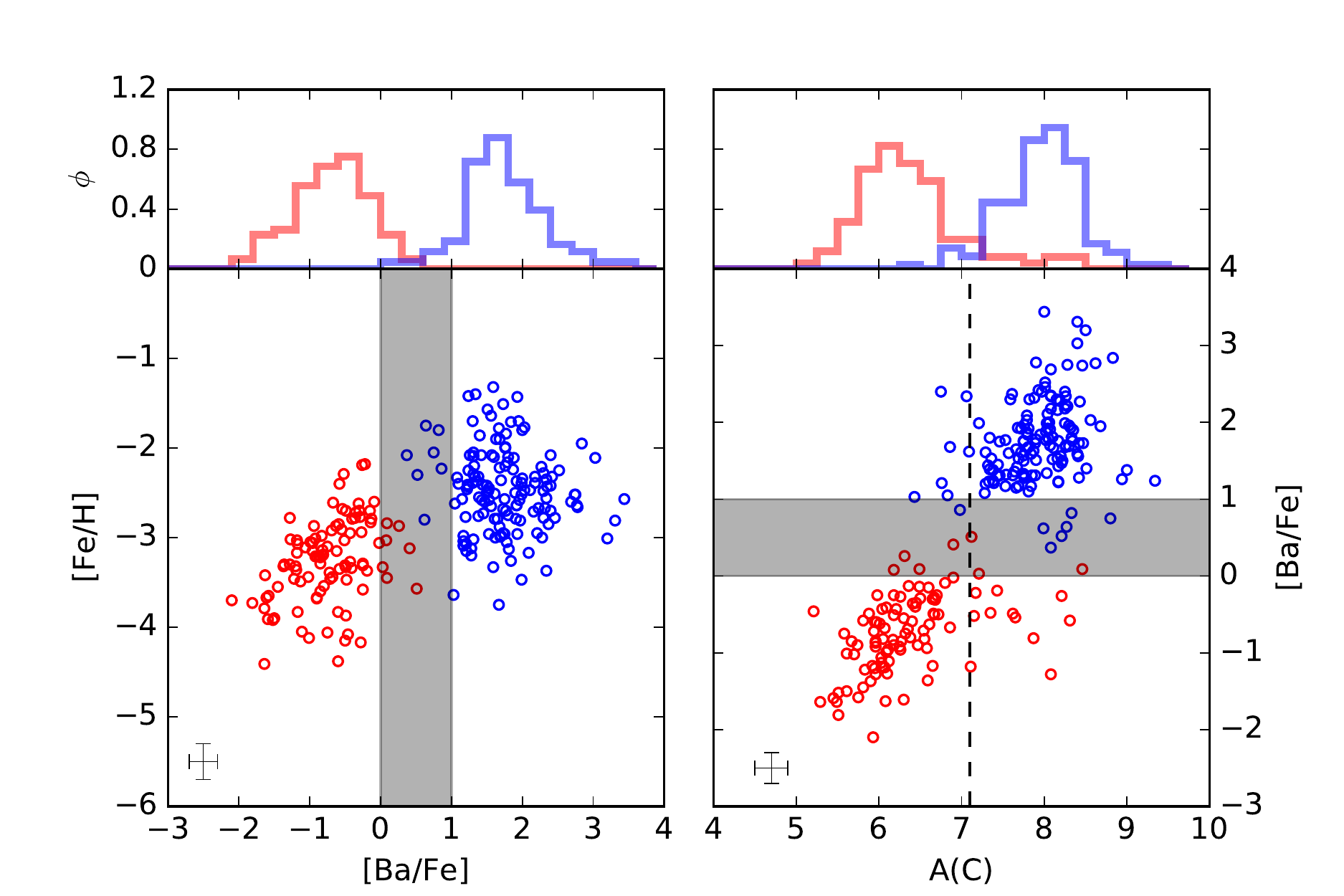}
\caption{The distribution of [Fe/H], as a function of [Ba/Fe] (left
panels), and [Ba/Fe], as a function of $A$(C) (right panels). As before,
the blue and red colors indicate the CEMP-$s/rs$ and CEMP-no stars,
respectively. The gray shaded region in both panels represents stars
with intermediate [Ba/Fe] ratios, 0.0 $<$ [Ba/Fe] $< +1.0$, which
complicates their sub-classification. The dashed vertical line in the
right panel is a fiducial at $A$(C) $= 7.1$, placed between the peaks of
the histograms of the $A$(C) distributions for Group~I and Group~II and
III stars shown in Figure~\ref{fscatter}. There are 15 stars with upper
limits on [Ba/Fe] $<$ 0.0 not shown in this figure. A typical
observational error bar is shown in the bottom left of each panel.
\label{fba}}
\end{center}
\end{figure*}

There are a number of interesting exceptions to the rule found among the
stars in our sample. For example, there are several CEMP-$s/rs$ stars
located in the low-C region in Figure~\ref{fscatter} (or in the
transition region between the two regions), whose binary status are
unfortunately unknown. Accounting for their low $A$(C), coupled with high
[Ba/Fe] ratios, remains a challenge. In addition, the five apparently single
CEMP-s$/rs$ stars in the high-C region of Figure~\ref{fbinarity} 
should yield interesting constraints on the source of their carbon and
barium enrichment.

While the great majority of CEMP-no stars with known binary status are
single stars, there are five CEMP-no stars in our sample that {\it are}
recognized binaries. Among these, three of the CEMP-no binaries are
located in the high-C region, while two CEMP-no binaries fall in the
low-C region. As mentioned in Section~\ref{results} above, there are
also several CEMP-no stars located in the high-C region but without
known binary status. It is indeed crucial to conduct further
radial-velocity monitoring of these exceptional stars, as well as to
determine their full elemental-abundance patterns, in order to better
constrain their origin.

\subsection{Interesting Outliers}\label{interestingstars}

We have noted a number of stars in our sample that can be considered
outliers, in the sense that they deviate in some manner from the bulk of
stars with similar sub-classifications and/or binary status. Exploration
of the numerous and varied possibilities to account for the underlying
causes of the recognized deviant cases is well beyond the scope of this
paper. However, for convenience, and to spawn their future detailed
study, Table~\ref{tbl-3} lists the recognized outliers, and their
properties -- [Fe/H], $A$(C), [C/Fe]$_c$, [Ba/Fe], [Eu/Fe],
classification and binarity status (1 indicates known single stars, 2
indicates known binaries) -- grouped along with stars of similar
behavior, as well as stars without known binary status but with
disparate $A$(C) and [Ba/Fe].

\subsection{Is A(C) a more Fundamental Indicator of CEMP Classification?}\label{acindicator}

Although the CEMP stars exhibit complex behaviors in the $A$(C)-[Fe/H]
space, there is indeed a clear separation between the CEMP-$s/rs$ and
CEMP-no stars, as seen in the marginal histogram of $A$(C) shown in
Figure~\ref{fscatter}. This separation is observed in Figure~\ref{fba}
as well. The left panels of Figure~\ref{fba} show the distribution of
[Fe/H] and [Ba/Fe], along with the marginal histogram of [Ba/Fe] 
for both the CEMP-$r/rs$ and CEMP-no stars, plotted
at the top. The shaded region with 0.0 $<$ [Ba/Fe] $<$ +1.0 was not
considered in the original definition of the two sub-classes
\citep{beers2005}, and later studies adopted different criteria for the
sub-classification of stars in this region. 

The right panels of Figure~\ref{fba} show the relation between [Ba/Fe]
and $A$(C), along with the marginal histogram of $A$(C) plotted at the
top. There exists a clear correlation between $A$(C) and [Ba/Fe]. The
vertical line represents a fiducial at $A$(C) = 7.1, placed at the
midpoint of the two $A$(C) peaks. Although there exists an overlap
between the stars sub-classified on the basis of [Ba/Fe], there is a
clear separation between the majority of stars in the two sub-classes. 

Among the interesting outliers described in the previous subsection (and
shown in Figure~\ref{fbinarity}), three of the eight CEMP-no stars with
$A$(C) $> 7.1$ are known binaries, while five are known single stars,
double the expected binary fraction based on that of most CEMP-no stars
according to \citet{hansen2016a}; further speculation is not yet warranted,
given the large Poisson errors in this fraction. The apparently
disparate stars shown in the right panel of Figure~\ref{fba} are of
interest as well. There is only one known binary among the CEMP-s/rs
stars in the upper left quadrant of the right panel of this figure
($A$(C) $\leq$ 7.1 and [Ba/Fe] $ > +$1.0), while three of the nine CEMP-no
stars in the lower-right quadrant ($A$(C) $ > $ 7.1 and [Ba/Fe] $< 0$)
are known binaries. Future radial-velocity monitoring of these
exceptional cases is clearly of interest.

The original definition of the CEMP sub-classes was purely empirical,
based on a dichotomy noticed in the distribution of [Ba/Fe] as a
function of [Fe/H] \citep[e.g.,][]{aoki2002,ryan2005}.
\citet{beers2005} quantified this dichotomy, which has been widely
adopted since. However, barium abundance measurements can only be
readily obtained from moderately-high to high-resolution spectroscopy,
which for fainter stars ($V > 14$) generally requires large telescope
time. In contrast, C-abundance estimates can be obtained from low- to
medium-resolution spectroscopy, which can be readily acquired by smaller
telescopes, such as the 2.5m telescope used for the Sloan Digital Sky
Survey \citep[SDSS,][]{york2000,lee2013}. The clear separation between
the CEMP-$s/rs$ and CEMP-no stars based on $A$(C) we have noted in this
study can thus greatly expedite the process of sub-classification. 

In addition, the degree of enhancement in carbon abundance is likely to
depend primarily on the nature (and mass) of the progenitors (although
dilution in the parent mini-halo may also play an important role); large
enhancements of $A$(C) by mass transfer from low- to
intermediate-mass AGB stars and lower enhancements in the natal clouds
of high-mass progenitors. Barium enhancement can arise from a
variety of processes, e.g., the $r$-process at low metallicity (from
progenitors of a currently uncertain nature; \citealt{roederer2014b}), and
the weak $s$-process in massive stars \citep{frischknecht2016}, as well
as from AGB nucleosynthesis \citep{bisterzo2011}. Therefore, we suggest
that the absolute carbon abundance, $A$(C), may be a more fundamental
criterion for separating the CEMP-$s/rs$ stars from CEMP-no stars. 

\begin{figure*}[tb!]
\includegraphics[scale=0.65]{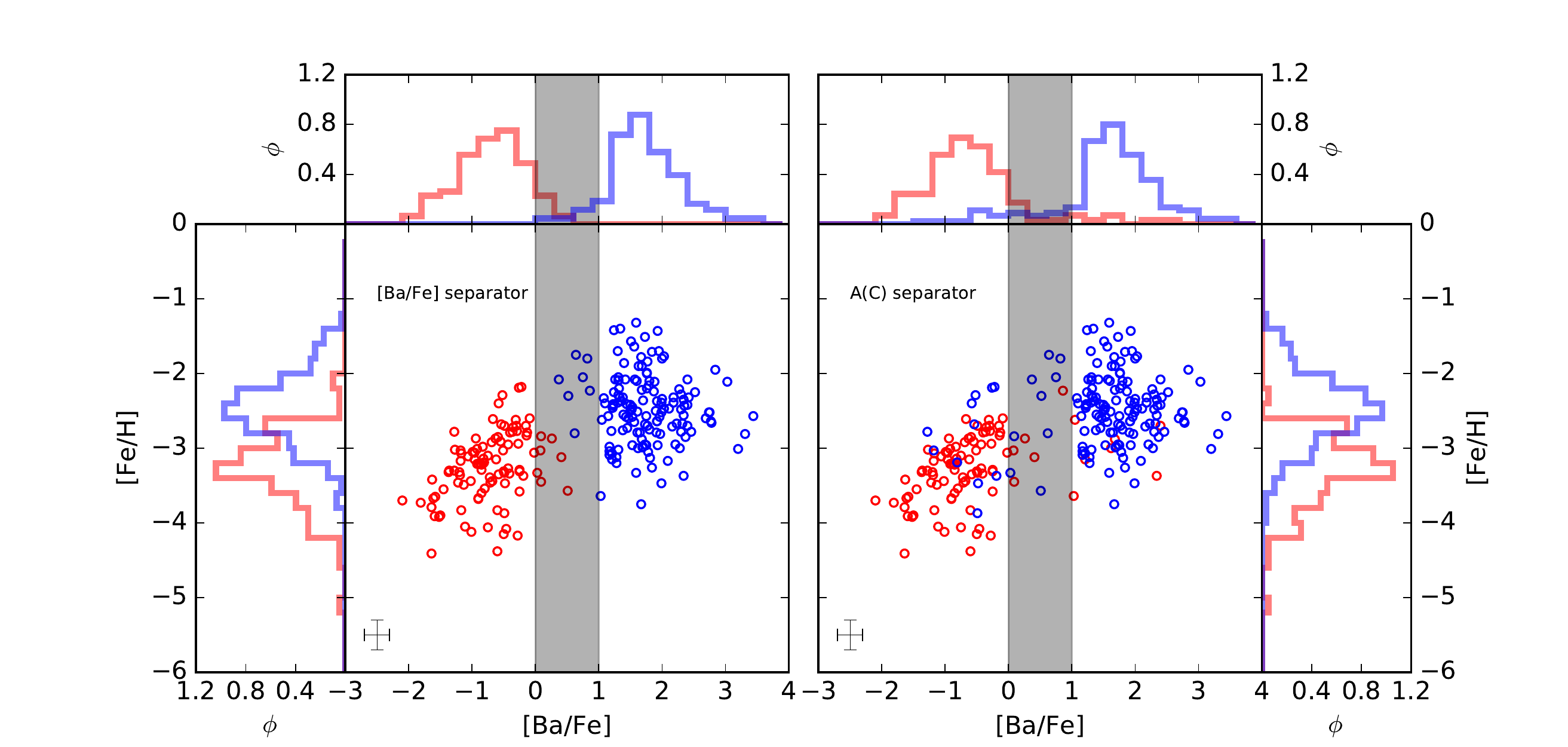}
\caption{Comparison between stars sub-classified into the CEMP-$s/rs$ and
CEMP-no categories using either the [Ba/Fe] or the $A$(C) criterion. As
before, the blue and red colors indicate the CEMP-$s/rs$ and CEMP-no
stars, respectively. The left panels correspond to use of the [Ba/Fe]
criterion, while the right panels are based on the $A$(C) criterion.
That is, CEMP-no stars are classified as such if they have $A$(C) $\leq
7.1$, while CEMP-$s/rs$ stars are classified as such if they have
$A$(C) $ > 7.1$. There are 15 stars with upper limits on [Ba/Fe] $<$
0.0 not shown in this figure. A typical observational error bar is shown
in the bottom left of each panel.
\label{fcomparison}}
\end{figure*}

In order to assess the efficacy of sub-classification for CEMP stars
based on $A$(C), we have re-classified the stars in our sample -- stars
with $A$(C) $\leq$ 7.1 are classified as CEMP-no stars and those with
$A$(C) $ > 7.1$ are considered CEMP-$s/rs$ stars. The left-hand panels
of Figure~\ref{fcomparison} are obtained from application of the
conventional classification criterion, based on [Ba/Fe]
\citep{beers2005}. The right-hand panels show the re-classified CEMP
stars, obtained from application of the suggested $A$(C) criterion. Both
classifications are listed in column (10) of Table~\ref{tbl-1}.

As can be verified from inspection of Figure~\ref{fcomparison}, the
distinctive separation of these two sub-classes of CEMP stars based on
$A$(C) is apparently {\it as effective as} that obtained from the
application of the conventional [Ba/Fe] criterion. Employing the $A$(C)
criterion, our sample of stars includes 159 CEMP-$s/rs$ stars and 115
CEMP-$no$ stars. Of the 159 re-classified CEMP-$s/rs$ stars, 20 were
originally classified as CEMP-$no$ stars, a ``success rate'' (if the
conventional classification approach is taken as ground truth) of 87\%.
For the 115 re-classified CEMP-no stars, 8 stars were classified as
CEMP-$s/rs$ stars by the traditional criterion, a success rate of 93\%.
The binary fractions of the re-classified CEMP-$s/rs$ and CEMP-$no$
stars changed from 86$\pm$21\% (30 of 35) to 76$\pm$18\% (32 of 42) and 
23$\pm$11\% (5 of 22) to 20$\pm$13\% (3 of 15), respectively, which are
not significant, given the large Poisson errors. We have applied the
$A$(C) criterion to the 31 unclassified CEMP stars listed at the end of
Table~\ref{tbl-1} -- those either without reported [Ba/Fe] abundance
ratios or located in the ambiguous zone (0.0 $<$ [Ba/Fe] $< $+1.0).
There are an additional 19 CEMP-no stars and 12 CEMP-$s/rs$ stars based
on the new classification scheme.\\\\

\section{Conclusions}\label{conclusion}

We have investigated the absolute carbon-abundance distribution of CEMP
stars based on a reasonably complete compilation of available
high-resolution spectroscopic data. The $A$(C) distribution of the CEMP
stars clearly exhibits (at least) a bimodality, as has been noted by a
number of previous authors. However, there exist complex behaviors
embedded in the $A$(C)-[Fe/H] space, not easily captured by description
as plateaus or bands; we suggest use of the terms high-C and low-C
regions. We separate CEMP stars into three groups -- Group~I, comprising
primarily CEMP-$s/rs$ stars, and Group~II and Group~III comprising
CEMP-no stars. Along with the apparent dichotomy in the absolute
abundance distribution of Na and Mg as a function of $A$(C) for the
CEMP-no stars, we suggest this provides the first clear observational
evidence for the existence of multiple progenitor populations of the
CEMP-no stars in the early Universe.

Based on the known binary status for a subset of the CEMP stars, we
strongly support the hypothesis that the carbon enhancement of the
CEMP-$s/rs$ stars is extrinsic, and due to mass-transfer of material
enriched by an AGB companion, while the carbon enhancement of CEMP-no
stars is intrinsic, and due to enrichment of their natal clouds by
high-mass progenitor stars. According to this view, the CEMP-no stars
are bona-fide second generation stars, as supported by other lines of
evidence (see summary in \citealt{hansen2016a}). 

We have identified a number of interesting outliers, worthy of further
exploration, that differ from the general behaviors of otherwise similar
stars in either their binary status or with disparate $A$(C) and [Ba/Fe]
ratios. 

Finally, we have presented evidence that the separation of CEMP-$s/rs$
stars from CEMP-no stars can be accomplished as well (or better) 
using the simple criterion $A$(C) $> 7.1$ for CEMP-$s/rs$ stars and  
$A$(C) $\leq 7.1$ for CEMP-no stars.  As $A$(C) can be obtained from low- to
medium-resolution spectroscopy, rather than the high-resolution
spectroscopy required for the former criterion based on [Ba/Fe] (and
[Ba/Eu]) ratios, this provides an efficient means to quickly isolate
the most interesting CEMP sub-classes in massive spectroscopic surveys
now and in the future.  Given the multiple nucleosynthetic pathways for the
production of Ba in the early Universe known (or suggested) to exist, we
assert that the $A$(C) criterion may also be more astrophysically fundamental.

We plan to employ the $A$(C) sub-classification approach to the
thousands of CEMP stars presently identified by SDSS/SEGUE and other
large surveys from the past, e.g., the HK survey \citep{beers1985,
beers1992} and the Hamburg/ESO survey \citep{christlieb2003}, enabling
consideration of potential differences in the spatial and kinematic
distributions of CEMP-$s/rs$ and CEMP-no stars (e.g.,
\citealt{carollo2014}), as well as in their relative frequencies as a
function of [Fe/H] (e.g., \citealt{lee2013}). Comparison of these
observables with the predictions of modern Galactic chemical evolution
models (e.g., \citealt{cote2016,crosby2016}) should prove illuminating.

Future progress requires a significant increase in the numbers of
CEMP-$s/rs$ stars and CEMP-no stars with available high-resolution
spectroscopy, so that their full elemental-abundance distributions can
be considered in more detail (both from the ground and in the near-UV
from space), and used to better constrain their likely progenitors. This
goal is being actively pursued. Additional long-term
radial-velocity monitoring of CEMP stars is also likely to pay
substantial scientific dividends. For both reasons, the identification
of, in particular, bright CEMP stars is being given high priority in our
ongoing survey efforts.   

There is also a clear need for additional development of theory and
modeling, to obtain deeper understanding of the nucleosynthesis
processes in operation in the suggested progenitors of CEMP stars. Links
between the CEMP phenomenon and the early-Universe initial mass function
have been considered by a number of previous authors (e.g.,
\citealt{lucatello2005b,tumlinson2007b,tumlinson2007a,suda2013,carollo2014,lee2014}), 
based on the relative frequencies of CEMP-$s/rs$ and
CEMP-no stars in the halo system. Here we have shown that this idea
might be extended by taking into account the presumed
differences in the masses of the progenitors of the Group~II and
Group~III CEMP-no stars.   

\acknowledgments
The authors thank C. Abate, W. Aoki, A. Maeder, G. Meynet, M. Pignatari,
R. Stancliffe, N. Tominaga, and K. Venn for useful discussions, as 
well as the anonymous referee for providing useful comments on the manuscript, 
which improved this paper. J.Y.,
T.C.B., V.M.P., K.C.R., D.C., and I.U.R. acknowledge partial support
from grant PHY 14-30152; Physics Frontier Center/JINA Center for the
Evolution of the Elements (JINA-CEE), awarded by the US National Science
Foundation. S.H. participated this work as an REU student at the
University of Notre Dame. T.T.H. acknowledges partial support from grant
AST-1108811, awarded by the US National Science Foundation.

This research made use of NASA's Astrophysics Data System, the
SIMBAD astronomical database, operated at CDS, Strasbourg, France, and
the SAGA database \citep{suda2008,suda2011,yamada2013}
(http://sagadatabase.jp). This work also made extensive
use of python.org, astropy.org, and scikit-learn.org.

\bibliography{bibliography}

%\clearpage

\cleardoublepage

\begin{deluxetable*}{lhhcrRRrrRRCcCr}
\tabletypesize{\tiny}
\tablewidth{0pt} 
\tablecaption{Literature Data and Classifications for Carbon-Enhanced
Metal-Poor (CEMP) Stars\label{tbl-1}}
\tablehead{
\colhead{Name} & \nocolhead{RA} & \nocolhead{DEC}& \colhead{T$_{\rm eff}$}& \colhead{$\log g$}&\colhead{ [Fe/H]} 
& \colhead{[C/Fe]} &\colhead{[C/Fe]$_{c}$} &\colhead{$A$(C)}&\colhead{[Ba/Fe]}
&\colhead{[Eu/Fe]}&\colhead{Class} &\colhead{Binarity\tablenotemark{a}}&\colhead{Interesting}&\colhead{Reference}\\
\colhead{} & \nocolhead{hms} & \nocolhead{dms}& \colhead{(K)}& \colhead{(cgs)}&\colhead{} 
& \colhead{} &\colhead{} &\colhead{(corrected)}&\colhead{}
&\colhead{}&\colhead{(Ba$\mid A$(C))} &\colhead{} &\colhead{outliers}&\colhead{}}  
\colnumbers
\startdata
\multicolumn{15}{c}{Group~I Stars}\\
\hline
                 52972-1213-507  &    09:18:49.9177  &   +37:44:26.8122  &  6463  &  4.34  &     -2.98  &      2.81  &      2.81  &      8.26  &                 1.7  &             \nodata  &       (~~\rm{s}\mid ~~\rm{s})  &                  \nodata  &        &            \citet{yong2013} \\ 
                     BD+04 2466  &    11:26:49.1912  &   +03:51:51.7735  &  5100  &   1.8  &      -1.9  &      1.24  &      1.34  &      7.87  &                1.63  &             \nodata  &       (~~\rm{s}\mid ~~\rm{s})  &                  \nodata  &        &         \citet{pereira2009} \\ 
                   BS 16077-077  &    11:55:16.5464  &   +28:21:00.9456  &  5900  &  3.19  &     -2.05  &      2.39  &      2.42  &      8.80  &                0.75  &                0.04  &       (~~\rm{s}\mid ~~\rm{s})  &                  \nodata  &        &           \citet{allen2012} \\ 
                   BS 16080-175  &    16:51:08.8187  &   +57:12:27.6286  &  6240  &   3.7  &     -1.86  &      1.94  &      1.94  &      8.51  &                1.40  &               -0.44  &       (~~\rm{s}\mid ~~\rm{s})  &                  \nodata  &        &           \citet{allen2012} \\ 
                   BS 17436-058  &    13:36:46.0331  &   +45:03:08.5395  &  5430  &   2.2  &      -1.9  &      1.71  &      1.73  &      8.26  &                1.68  &                1.28  &       (\rm{r/s}\mid \rm{r/s})  &                  \nodata  &        &           \citet{allen2012} \\ 
                     CD-62 1346  &    21:06:02.9161  &   -61:33:44.6845  &  5300  &  1.70  &     -1.57  &      0.93  &      1.04  &      7.90  &                1.51  &             \nodata  &       (~~\rm{s}\mid ~~\rm{s})  &                  \nodata  &        &         \citet{pereira2012} \\
\enddata
\end{deluxetable*}

\cleardoublepage
\begin{deluxetable*}{lRRRRRRl}
\tabletypesize{\scriptsize}
\tablecaption{Na and Mg abundances for the Group~II and Group~III CEMP-no stars\label{tbl-2}}
\tablewidth{0pt}
\tablehead{
\colhead{Name}&\colhead{[Fe/H]}& \colhead{$A$(C)} &\colhead{[Na/Fe]}& \colhead{$A$(Na)}& \colhead{[Mg/Fe]} &\colhead{$A$(Mg)} & \colhead{Reference}} 	
\startdata 
\hline
\multicolumn{8}{c}{Group~II CEMP-no Stars} \\
\hline
BD+44 493 & -3.83 & 5.95 & 0.30 & 2.71 & 0.46 & 4.23 & \citet{ito2013} \\
BD-18 5550 & -3.2 & 6.05 & \nodata & \nodata & 0.49 & 4.89 & \citet{roederer2014} \\
BS 16929-005 & -3.34 & 6.09 & 0.03 & 2.93 & 0.30 & 4.56 & \citet{yong2013} \\
CD-24 17504 & -3.41 & 6.12 & -0.23 & 2.60 & 0.34 & 4.53 & \citet{jacobson2015} \\
CS 22185-007 & -3.06 & 6.11 & -0.02 & 3.16 & 0.49 & 5.03 & \citet{roederer2014} \\
CS 22189-009 & -3.92 & 5.51 & -0.10 & 2.22 & 0.44 & 4.12 & \citet{roederer2014} \\
CS 22873-128 & -3.32 & 5.90 & -0.11 & 2.81 & 0.66 & 4.94 & \citet{roederer2014} \\
\enddata
\tablecomments{$A$(C) is based on the corrected carbon abundance ratio,
[C/Fe]$_c$; the listed references are the same as in Table~\ref{tbl-1}.
$A$(Na) and $A$(Mg) are obtained based on the
Solar abundances of Na and Mg from \citet{asplund2009}.}
\tablenotetext{a}{This star has only reported $A$(Na) and $A$(Mg).  [Na/Fe] and [Mg/Fe] were calculated based on the \citet{asplund2009} Solar abundance.}
\tablenotetext{b}{This star has only reported [Na/H] $< -5.7$ and [Mg/H]
= $-4.45$.} 
\tablenotetext{c}{[Mg/Fe] was calculated based on the reported [Mg/H]$-$[Fe/H].}
\end{deluxetable*}

\cleardoublepage
\begin{deluxetable*}{lhRrrRRCcl}
\tabletypesize{\scriptsize}
\tablecaption{Interesting Outliers among the CEMP Stars\label{tbl-3}}
%\tablewidth{0pt}
\tablehead{
\colhead{Name}& \nocolhead{$\log g$} &\colhead{[Fe/H]}&\colhead{$A(C)$}& \colhead{[C/Fe]$_{c}$}&\colhead{[Ba/Fe]} &\colhead{[Eu/Fe]} &\colhead{Class}& \colhead{Binarity} &\colhead{Reference}\\
\colhead{}& \nocolhead{(cgs)} &\colhead{}&\colhead{(corrected)}& \colhead{}&\colhead{} &\colhead{} &\colhead{(Ba$\mid A(C)$)} &\colhead{} & \colhead{}
} 	
\startdata 
\multicolumn{10}{c}{Single CEMP-no Stars with high $A$(C), low [Ba/Fe]} \\
\hline
                    CS 29498-043   &  0.50 &  -3.87 & 7.62       & 3.06      & -0.49        &  <0.23  &      (~\rm{no}\mid ~~~\rm{s})  &        1   &    \citet{roederer2014}\\ 
                    HE 0405-0526   &  3.86 &  -2.18 &   7.17    &   0.92     &    -0.22   &\nodata      &      (~\rm{no}\mid ~~~\rm{s})  &         1  &    \citet{hansen2016a} \\ 
                    HE 1133-0555   &  1.31 &  -2.40 & 8.31       & 2.28       &  -0.58       &  \nodata  &      (~\rm{no}\mid ~~~\rm{s})  &         1  &    \citet{hansen2016a} \\ 
                    HE 1302-0954   &  2.40 &  -2.25 &    7.37    &   1.19    & <-0.53    & \nodata    &      (~\rm{no}\mid ~~~\rm{s})  &         1  &    \citet{hansen2016a} \\ 
                      HE 1410+0213  &  2.00 &  -2.19 &   8.21     & 1.97      &     -0.26  & \nodata   &      (~\rm{no}\mid ~~~\rm{s})  &         1  &       \citet{cohen2013}\\ 
\hline
\multicolumn{10}{c}{Binary CEMP-no Stars with high $A$(C), low [Ba/Fe]} \\
\hline
                    CS 22957-027   &  2.65 &  -3.19 & 7.87       & 2.63      & -0.81       &   <1.07 &      (~\rm{no}\mid ~~~\rm{s})  &         2  &    \citet{roederer2014}\\
\enddata
\tablecomments{The references for the claimed binarity are the same as in Table~\ref{tbl-1}.}
\tablenotetext{a}{CS 22964-161A has a higher reported $A$(C) (7.64)
than CS 22964-161B, by 0.2 dex, and is not listed in this table.}
\end{deluxetable*}

\end{document}